\documentclass[times]{article}

\usepackage{cite}
\usepackage[dvips]{graphicx}
\usepackage[cmex10]{amsmath}
\usepackage{psfrag}
\usepackage{bm}
\usepackage{comment}

\def \bmx {\bm{x}}
\def \bmy {\bm{y}}
\def \bmnu {\bm{\nu}}
\def \bmn {\bm{n}}

\def \bmH {\bm{H}}

\def \EE  {\mathsf{E}}

\def \llangleBigg {\Bigg\langle\kern-4.2pt\Bigg\langle}
\def \rrangleBigg {\Bigg\rangle\kern-4.2pt\Bigg\rangle}

\def \argmax {\mathop{\rm argmax}}

\title{Performance Analysis of Signal Detection 
using Quantized Received Signals of Linear Vector Channel}
\author{Kazutaka NAKAMURA \thanks{
This work was supported by the Grant-in-Aid for Scientific Research on Priority Areas (No.~18079010), 
the Ministry of Education, Culture, Sports, Science and Technology, Japan.
This paper was presented in part at the IEICE Transactions on Fundamentals of
Electronics, Communications and Computers (Japanese Edition) Vol. J91-A, No. 5,
pp. 550--558, 2008.}}
\date{Department of Systems Science, Graduate School of
Informatics, Kyoto University\\
36-1, Yoshida-Honmachi, Sakyo-ku, Kyoto, 606-8501, Japan\\
E-mail: knakamur@i.kyoto-u.ac.jp}

\begin{document}

\maketitle
\sloppy
\begin{abstract}
Performance analysis of optimal signal detection 
using quantized received signals of
a linear vector channel, which is an extension of 
code-division multiple-access (CDMA) 
or multiple-input multiple-output (MIMO) channels, 
in the large system limit, is presented in this paper. 
Here the dimensions of channel input and
output are both sent to infinity while their ratio remains fixed.
An optimal detector is one that uses a true channel model, 
true distribution of input signals, and perfect knowledge about quantization. 
Applying replica method developed in statistical mechanics, 
we show that, in the case of a noiseless channel, 
the optimal detector has perfect detection ability under certain conditions,
and that for a noisy channel 
its detection ability  decreases monotonically 
as the quantization step size increases. 
\end{abstract}

\section{INTRODUCTION}
In modern wireless communication systems, many processes are 
performed on digital signals converted by an analog-to-digital (A/D) converter.
In code-division multiple-access (CDMA) system which is one of key
technologies for third generation cellular phone system, 
spreading, de-spreading, and code acquisition are all performed on digital
signals.

Although there have been many studies on 
performance analysis of digital communication systems,
information-theoretic performance analyses usually assume that
the received analog signals are
available~\cite{Verdu_Book_1998,TTanaka_CDMA_IEEE,GV_IT_2005}.
Previous studies~\cite{IH_2000,SMP_2005}, considering the quantized
effect of received signals, used  
non optimal signal detection schemes, such as linear filters.

In this paper, we evaluate the performance of optimal signal detection 
using the quantized received signals of a linear vector channel, which 
is an extension of CDMA or multiple-input multiple-output (MIMO) channels, 
in the large system limit, where the dimensions of channel input and output are
both sent to infinity while their ratio remains fixed.

\section{CHANNEL MODEL AND QUANTIZATION OF RECEIVED SIGNALS}
We consider a $K$-input, $N$-output linear vector channel, defined as follows.
Let $\bmx_0=(x_{01},\,\ldots,\,x_{0K})^T \in \{+1,\,-1\}^K$ denote the
channel input vector, and $\bmy=(y_1,\,\ldots,\,y_N)^T$ denote
the output vector, given a linear transform $H\bm{x}_0$ of the inputs, 
where $H$ is an $N \times K$ channel matrix.  
Assuming the additive white Gaussian (AWGN) channel (variance $\sigma_0^2$), 
the outputs are described as 
\begin{align}
\bmy = H \bmx_0 + \sigma_0 \bmnu,
\end{align}
where the components of $\bmnu=(\nu_1,\,\ldots,\,\nu_N)^T$ follow 
the normal distribution $\mathcal{N}(0, \,1)$. 
To simplify the analysis, we assume a random channel matrix;
the elements $\{H_{\mu k}\}$ are independent and identically distributed
(i.i.d.) with mean zero and variance $1/N$ .

In this study, we consider signal detection using received
signals, which are quantized by an A/D converter at the receiver.
We assume an A/D converter that quantizes received
signal $y$ into integer $n$, which satisfies the condition 
$(n-1/2)d < y < (n + 1/2)d $, where $d>0$ is the quantization step size.
The probability distribution of the quantized received signals
$\bmn=(n_1,\ldots,\,n_N)^T$, given the input $\bmx_0$ and channel matrix
$H$, is represented as 
\begin{align}
&P_0(\bm{n}|H \bmx_0) = \prod_{\mu=1}^N \rho_0(n_\mu|\bm{H}_\mu^T
\bmx_0), \\ 
&\rho_0 ( n_\mu | \bm{H}_\mu^T \bmx_0) \nonumber \\
&=
\frac{1}{\sqrt{2 \pi \sigma_0^2}}
\int_{(n_\mu -\frac{1}{2})d}^{(n_\mu +\frac{1}{2})d}
\exp \left[
-\frac{ (y - \bm{H}_\mu^T \bmx_0)^2}
{2\sigma_0^2}
\right] dy
\nonumber \\
&=
Q\! \left[\frac{\! \bigl(n_\mu -\frac{1}{2}\bigr)d - \bm{H}_\mu^T
\bmx_0 \!}{\sigma_0}\right]
\! - \!
Q\! \left[\! \frac{\bigl(n_\mu +\frac{1}{2}\bigr)d - \bm{H}_\mu^T
\bmx_0 \!}{\sigma_0}\right],
\label{eq:channel_quantum}
\end{align}
where $\bm{H}_\mu^T$, which denotes the $\mu$th row of $H$ and   
$Q(x) =  \int_x^\infty Dt$, $Dt=(2\pi)^{-1/2}e^{-t^2/2}\,dt$.
Equation \eqref{eq:channel_quantum} represents the
input($\bmx_0$)-output($\bmn$) characteristic of the quantization
channel.

The signal detection of channel input $\bmx_0$, given the quantized
received signals $\bmn$ and the channel matrix $H$, can be solved by  
an inference scheme based on Baysian inference.
We define a prior as $P_0(\bmx_0)$, and assume $P_0(\bmx_0) = 1/2^K$.
The detector assumes the channel model of quantization channel to be 
$P(\bmn|H\bmx) = \prod_{\mu=1}^N \rho(n_\mu|\bmH_\mu \bmx)$ 
and the prior distribution to be $P(\bmx)$.
The posterior distribution is represented as 
\begin{align}
P(\bmx|\bmn,\,H)=
\frac{ P(\bmn |H\bmx) P(\bmx)}
{\sum_{\bmx} P(\bm{n}|H\bmx) P(\bmx)}.
\label{eq:Bayes_post_prob}
\end{align}
Maximizer of the Posterior Marginals (MPM) estimation
$\hat{x}_k = \argmax_{x_k} \sum_{x_{l \neq k}} P(\bmx|\bmn,\,H)$ is the
optimal inference scheme for minimizing the component-wise estimation
error probability, when the assumed channel model and prior distribution are 
matched to the true ones. 
Hereafter, we call the detector using true channel model and prior
distribution as an optimal detector.

\section{PERFORMANCE ANALYSIS USING REPLICA METHOD}
We evaluate the performance of the optimal detector using the quantized
received signals in the large system limit, where $K,~N \to \infty$ while the
ratio $\beta=K/N$ is kept finite.
This paper~\cite{TT_ALT_2004} studies the multiuser detection performance  
of a CDMA channel with an arbitrary memoryless channel and arbitrary 
distribution of channel inputs in the large system limit, using the replica
method. 
Since the defined quantized channel \eqref{eq:channel_quantum} is 
an example of memoryless channels, 
one can apply the analysis reported in \cite{TT_ALT_2004} 
to the performance evaluation of the optimal detector using quantized received
signals in the large system limit.
Although there is no rigorous justification for the replica method,
we assume the validity of the replica method, and related techniques throughout
this paper.

In the large system limit, the estimation error probability of the optimal
detector
\begin{align}
P_b &= \EE_{\bmx_0,\,\bmnu,\,H} \left[ \frac{1}{K}\sum_{k=1}^K
\frac{(1-\hat{x}_k x_{0k} )}{2} \right],
\label{eq:BER_def}
\end{align}
where $\EE_X[\cdots]$, which denotes the average with respect to $X$,
can be evaluated as 
\begin{align}
P_b = Q(\sqrt{E}).
\label{eq:BER_RS_solution}
\end{align}
The parameter $E$ is determined by solving the following equations for
$\{m,\,E\}$,
\begin{align}
m &= \int \tanh \left(E+\sqrt{E} \,z \right) Dz, 
\label{eq:RS_saddle_m} 
\\
E &= \sum_{n=-\infty}^\infty \int  
 \frac{[\overline{\rho}_0^\prime\left(n \big|\sqrt{\beta m}\,t\right)]^2}
 {\overline{\rho}_0 \left(n\big|\sqrt{\beta m}\,t\right)} Dt,
\label{eq:RS_saddle_E} 
\end{align}
where $\bar{\rho}_0(n|\sqrt{\beta m})$ is defined as
\begin{align}
&\overline{\rho}_0 \left( n \big| \sqrt{\beta m} \,t \right)
=Q\left[ \frac{ \left( n-\frac{1}{2} \right) d - \sqrt{\beta m}\,t}
{\sqrt{\sigma_0^2+ \beta (1-m)}} \right]
\nonumber \\
&~~~~~~~~~~-
Q \left[ \frac{ \left( n+\frac{1}{2} \right) d - \sqrt{\beta m}\,t}
{\sqrt{\sigma_0^2+ \beta (1-m)}} \right],
\label{eq:P_0_bar_true}
\end{align}
and $\bar{\rho}_0^\prime(n|\Delta) = \frac{\partial}{\partial
\Delta}\bar{\rho}_0(n|\Delta)$.

For some values of the parameters $(\sigma_0,\,d,\,\beta)$, 
equations~\eqref{eq:RS_saddle_m} and \eqref{eq:RS_saddle_E} 
have three solutions, and 
these solutions can be distinguished by the value of
estimation error probabilities $P_b$;
we define good, intermediate, 
and bad solutions from the smallest to the largest value of $P_b$.
Furthermore, we define the correct solution as
one minimizes the
function $\mathcal{F}$, which is defined as 
\begin{align}
&\mathcal{F} 
=
\frac{1}{\beta} \sum_{n=-\infty}^\infty
\! \int
\overline{\rho}_0
\left(
n \big| \sqrt{\beta m}\,t
\right)
\log
\overline{\rho}_0
\left(
n \big| \sqrt{\beta m}\,t
\right)  Dt
\nonumber \\
&- \frac{1}{2} EM  
+\int \log \left[ 2 \cosh \left( E + \sqrt{E} z \right) \right]~Dz.
\end{align}

It is known that the correct solution corresponds to the
performance of the optimal detector.
Since the exact evaluation of marginal posterior probability is usually
computationally infeasible, we have to resort to approximation techniques.
It is also known that a bad solution corresponds to the performance of
the belief propagation based on iterative algorithm~\cite{YK_2003}, which
produces approximate marginal posterior probabilities in low time complexity.

\section{RESULTS}
\subsection{NOISELESS CHANNEL}
\begin{figure}
\begin{center}
\includegraphics[angle=270,scale=0.20]{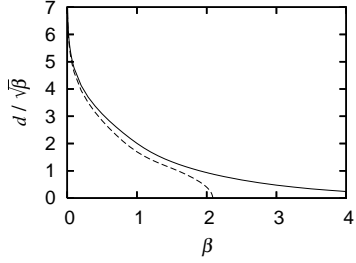}
\end{center}
\vspace{-5mm}
\caption{
Bifurcation diagram for the optimal detector over a noiseless channel.}
\label{fig:phase_zero_noise}
\end{figure}

\begin{figure}
\begin{center}
\includegraphics[angle=270,scale=0.20]{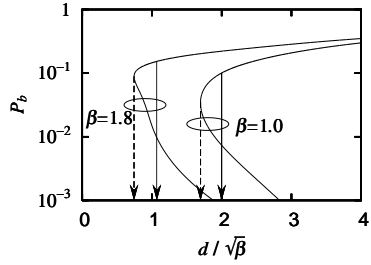}  
\end{center}
\vspace{-5mm}
\caption{Estimation error probabilities for
the optimal detector over a noiseless channel. 
Solid lines with arrows represent the upper bound of $d/\sqrt{\beta}$,
for which the optimal detector can achieve the perfect detection.
Broken lines with arrows represent the lower bounds of $d/\sqrt{\beta}$,
for which coexistence of the solutions occurs. }
\label{fig:BER_zero_noise}
\end{figure}

For a noiseless channel ($\sigma_0 = 0$), 
the equations~\eqref{eq:RS_saddle_m} and \eqref{eq:RS_saddle_E}
have the perfect detection solution $m=1$, $E=\infty$, which provides $P_b=0$
for any values of $d$ and $\beta$.
Figure~\ref{fig:phase_zero_noise} shows the bifurcation diagram.
The optimal detector can achieve the perfect detection
in the lower left side of the solid curve, 
and the coexistence of the solutions occurs above the broken curve.
Figure~\ref{fig:BER_zero_noise} shows the estimation error probabilities
of the optimal detector.

\subsection{NOISY CHANNEL}
\begin{figure}
\begin{center}
\includegraphics[angle=270,scale=0.20]{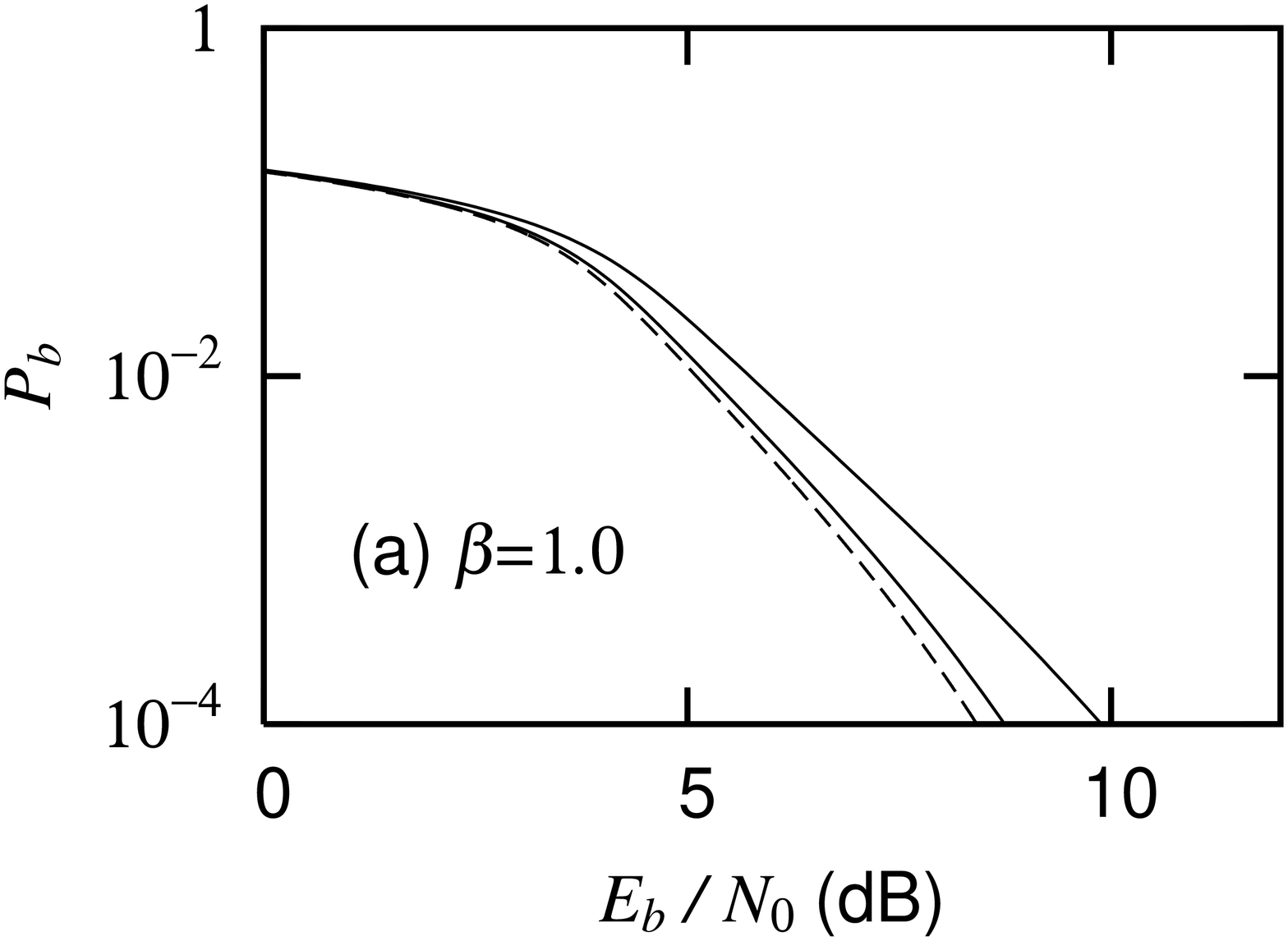}
\includegraphics[angle=270,scale=0.20]{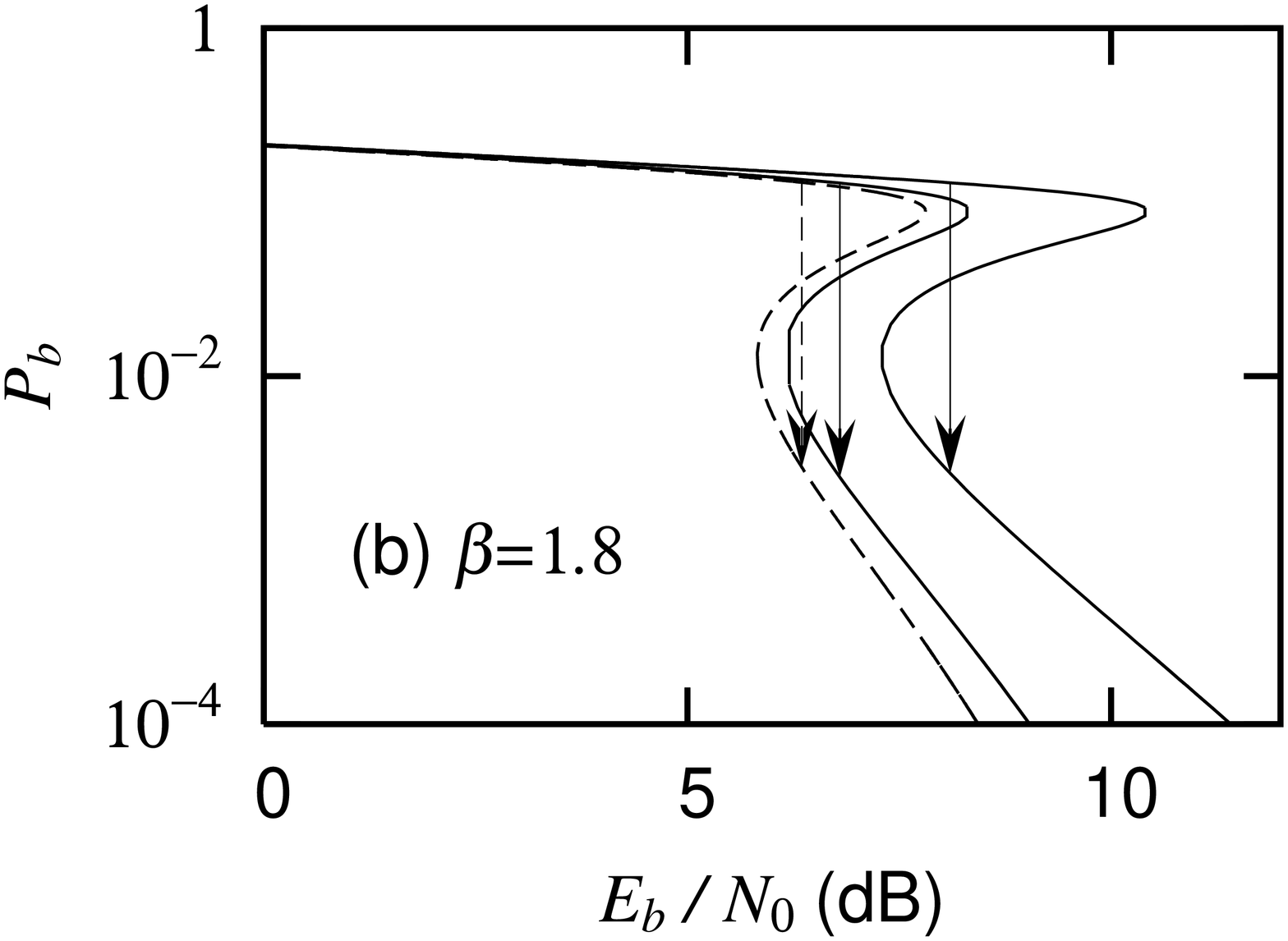}
\end{center}
\vspace{-5mm}
\caption{Estimation error probabilities versus $E_b/N_0$.
(a) $\beta=1.0$, (b)$\beta=1.8$.
The solid curves represent $d/\sqrt{\beta}=0.5$(right), $0.25$(left),
and the broken curve represents  $d/\sqrt{\beta}=0$.
The arrows represent the values of $E_b/N_0$, below which
the bad solution becomes correct solution. }
\label{fig:BER_vs_EbN0}
\end{figure}

For a noisy channel ($\sigma_0 > 0$), 
the optimal detector cannot exhibit perfect detection, 
due to the effect of quantization noise.
Figure~\ref{fig:BER_vs_EbN0} shows estimation error probabilities versus
$E_b/N_0 = 1/(2 \sigma_0^2)$.
The detection ability of the optimal detector decreases monotonically 
as the quantization step size increases. 


\section{CONCLUSION}
In this paper, we have evaluated the performance of an optimal detector using
the quantized received signals of a linear vector channel in the large system
limit. Applying the replica method, 
we have shown that, for a noiseless channel, under certain
conditions, the optimal detector has a perfect detection ability,
i.e., it is not affected by quantization noise. 
We have also shown that, for a noisy channel, 
the optimal detector's detection ability decreases monotonically
as the quantization step size increases.

\bibliographystyle{IEEEtran}
\bibliography{IEEEabrv,kn_isita08}

%

\end{document}